# Public Authorities as Defendants:
## Using Bayesian Networks to determine the Likelihood of Success for Negligence claims in the wake of Oakden

Scott McLachlan[1,2], Evangelia Kyrimi[1], Norman E Fenton[1]

[1] Risk and Information Management (RIM), Queen Mary University of London, UK

[2] Health informatics and Knowledge Engineering Research (HiKER) Group

**Abstract**

Several countries are currently investigating issues of neglect, poor quality care and abuse in the aged care sector. In most cases it is the State who license and monitor aged care providers, which frequently introduces a serious conflict of interest because the State also operate many of the facilities where our most vulnerable peoples are cared for. Where issues are raised with the standard of care being provided, the State are seen by many as a deep-pockets defendant and become the target of high-value lawsuits. This paper draws on cases and circumstances from one jurisdiction based on the English legal tradition, Australia, and proposes a Bayesian solution capable of determining probability for success for citizen plaintiffs who bring negligence claims against a public authority defendant. Use of a Bayesian network trained on case audit data shows that even when the plaintiff case meets all requirements for a successful negligence litigation, success is not often assured. Only in around one-fifth of these cases does the plaintiff succeed against a public authority as defendant.

**Introduction**

In late 2016 a review into the conduct of care at Oakden Older Persons Mental Health Service (OPMHS), hereafter referred to as Oakden, was commissioned[1]. Resulting from a complaint made by the wife of a patient who sustained unexplained injuries while an in-patient at the facility, the report's authors spent 12 weeks interviewing Oakden's staff and reviewing clinical documentation, care processes and procedures, and even assessing the facility itself. Their report was highly critical of both the facility and the standard of care provided to patients[2]. The reviewers were disturbed by the experience, having witnessed and reviewed patient care more akin to psychiatric hospitals of a previous era in an environment they described as *unable to embrace modern patterns of care*[3].

Media reports since release of both *The Oakden Report*[4] and the subsequent ICAC investigation's report *Oakden: A shameful chapter in South Australia's History*[5] suggest that families of Oakden patients are preparing lawsuits against the Government[6]. When multiple personal injury law firms, publicly admonished as *ambulance chasers*[7] for such practices, have come forward to suggest *significant compensation* may be achieved from negligence claims[8], families may be wondering *who they can sue*, *for what*, and *how much is significant compensation*?

---

[1] A Groves, D Thompson, D McKellar & N Proctor, 'The Oakden Report' (2017, SA Health, Department for Health and Ageing). Adelaide, South Australia, at 1.
[2] The Hon. Bruce Lander QC, 'Oakden: A shameful chapter in South Australia's History', (February 28, 2018, Independent Commissioner Against Corruption)
[3] 1 at 115.
[4] 1
[5] 2

We contend that a better question might be to ask how likely it is that a negligence lawsuit against a Government or Statutory body might result in such a financially promising outcome? It is this question that the remainder of this paper will seek to address.

**Oakden**

The Oakden and ICAC reports identified significant issues regarding the care and safety of patients housed in the facility. These included: (i) as many as 40 patient falls and other injuries every month[9]; (ii) excessive use of physical and chemical restraint as a method of patient management or control[10]; (iii) a monthly average of 10-15 medication errors[11]; and, (iv) a large number of sexual and physical assaults by staff against patients that had been reported to the police[12], the Office of the Public Advocate (OPA)[13], and the Office for Public Integrity (OPI)[14]. The ultimate outcome recommended for Oakden was closure[15], which has subsequently occurred. Staff and patients were distributed among a number of other South Australian aged care facilities.

It cannot be understated that, on appearance, there may be many causes of action that could be initiated against numerous defendants, be it: patient care attendants (PCAs); registered nurses (RNs); doctors; the triumvirate of doctors[16] who were seen by many as being in a position of responsibility for the facility; the State's chief psychiatrist who had multiple opportunities to step in and make his review prior to being required to do so; the Director or CEO of the local area health board; or State health Minister or federal Ministers for health or aged care. During the time that any particular aged person was in-patient at Oakden, any number of these actors may have been responsible for, or played some part, in the events that took place and which resulted in harm. However, the first consideration is going to be the type and degree of harm. It is not likely sufficient to simply say "my aged relative was in Oakden at the time, they *may have* been harmed". Or even "my aged relative was there and I am psychologically disturbed at what *might have* happened to them[17]". It is going to be those patients who were physically injured, or who died[18], in unexplained or insufficiently explained circumstances whose families are going to be the potential litigants in any such actions.

---

[6] Tom Fedorowytsch, 'Oakden victims' families push for justice, Premier Jay Weatherill vows to hold workers accountable', *ABC News* (online), 1 March 2018 <https://www.abc.net.au/news/2018-03-01/premier-promises-justice-for-families-of-oakden-patients/9497956>.
[7] Term adopted from: Fiona McLeay, 'Commissioner warns lawyers over marketing tactics', (Media Release, 13 July 2018) <https://www.lsbc.vic.gov.au/documents/Media_Release-%20Commissioner-warns-lawyers-over-marketing-tactics-2018.pdf>.
[8] Gerard Malouf and Partners, 'Did Oakden's aged care facility commit nursing home negligence?', (08 May 2018) <https://www.gerardmaloufpartners.com.au/Publication-2644-Did-Oakden-27s-aged-care-facility-commit-nursing-home-negligence.aspx>, and: Emma Mead, 'Mistreatment in Aged Care', (02 Oct 2018) <https://www.burkemeadlawyers.com.au/compensation-and-insurance-litigation/mistreatment-in-aged-care/>.
[9] 1 at 82.
[10] 1 at 81.
[11] 1 at 82.
[12] 1 at 70; and, 2 at 5.
[13] 1 at 82.
[14] 2 at 24 – 25; which details ten (10) separate complaints over four years of staff assaulting a patient physically and/or verbally, or arising from improperly performed or unconsented medical treatments.
[15] 1 at 115.
[16] 2 at 18.
[17] *Law Reform (Miscellaneous Provisions) Act* 1956 (NT), s 25: Liability for mental or nervous shock in respect of injury is extended to include a parent, spouse or de facto partner of someone who was *killed, injured or put in*

Many reports have been made to police yet few, if any, prosecutions have been commenced against staff members or those in positions of responsibility for Oakden. This is because it can be difficult to meet the standard required for a criminal prosecution when: (i) your victim, often your primary witness, may lack capacity[19]; (ii) there is a known culture of secrecy[20] that suppressed reporting[21] so that all such matters were kept *in-house*[22] and out of the public eye; and, (iii) what patient and other records there are can be more readily framed as mere fiction than an accurate reporting of events and patient care[23]. There would be little disagreement that the facility and care staff owed a *duty of care*[24], and in the case of those specialist psychiatrists who undertook to set the standard for treatment and had oversight of the facility, that the higher *standard of care* may be considered appropriate[25]. However, the issues that restrain criminal prosecutions are compounded when one seeks to identify particular individuals as defendants who would need to be evidenced as having visited an actual harm on a specific patient, and who therefore should be accountable for damages resulting from that harm. An even greater difficulty facing patients and family wishing to sue aged care facilities and their staff for abuse is that the law presently limits damages for elderly people. In most cases, economic loss, or lost income, is usually the largest single damage claimed by most people. However, such a claim is not available to retirees, and certainly not where they are long-term in-patients or on a pension or benefit[26]. As such, they and their families are limited to suing for: (i) pain and suffering[27]; (ii) past and future medical expenses; and, (iii) past and future care.

**Government and Public Authorities as Defendant**

Across Australia, the Commonwealth is responsible for primary mental health care, with delivery supported through community-based primary mental health services provided by

---

*peril*, and only to other family members where the person was *killed, injured or put in peril within their sight or hearing*.

[18] Such as Graham Rollbusch, a dementia patient who was murdered in his sleep by another dementia patient who staff knew was violent and aggressive, and who had threatened and assaulted Mr Rollbusch on two previous occasions. See: Andrew Hough, 'Elderly patient at scandal-plagued Oakden aged care home killed in room just weeks after attacked assaulted him, inquest told' (2017, The Advertiser) <https://www.adelaidenow.com.au/news/south-australia/elderly-patient-at-scandalplagued-oakden-aged-care-home-killed-in-room-just-weeks-after-attacker-assaulted-him-inquest-told/news-story/1c9d75fbb99ef2247615c4ead55a75ff>.

[19] While the general position in Australia is a presumption that every person is competent to give evidence, unless proven otherwise, and even though being diagnosed with dementia does not necessarily mean that you are no longer competent, the patients in Oakden were there because they were at the most severe end of the scale for age-related psychiatric illness such that capacity and competence are likely to be a significant pre-trial and appeal factor.

[20] 2 at 18; and, 1 at 95.

[21] 1 at 97.

[22] 2 at 18.

[23] 1 at 78-79.

[24] To the standard which a reasonable person would expect in similar circumstances.

[25] While other jurisdictions have provisions defining the standard of care for professionals, the common law position of *Rogers v Whitaker* governs the standard of care for professionals in the Northern territory.

[26] *Amaca Pty Ltd v Latz, Latz v Amaca Pty Ltd* [2018] HCA 22 – the court considered and unanimously found that the loss of expected income from a Commonwealth aged pension or superannuation pension was not a compensable damage after a negligently caused premature death.

[27] Which in many states requires assessment of factors such as duration, and the resulting damages are limited by *percentage* or *points scales*. For example: awards for non-pecuniary damages in the Northern Territory are governed the *Personal Injuries (Liabilities and Damages) Act* 2003 (NT) ss 24-28, and are based on the degree of impairment - which may be difficult to establish for Oakden patients who, in order to have become Oakden patients, were already impaired by their disease to the severest degree.

general practitioners and private specialist providers (psychiatrists, psychologists, nurses and other allied health professionals) through programs usually under control of Primary Health Networks (PHNs)[28]. The Commonwealth also takes the lead role in providing for Aged Care services, funding the Residential Aged Care sector and many dementia-specific treatment programs both in the community and in specialist Residential Aged Care Facilities (RACF)[29]. In general, state-funded health services provided through each Local Health Network (LHN) are responsible for the day to day provision of specialised mental health services[30]. While most aged care facilities are corporately or privately owned and only certified or licensed by the government authority, Oakden was owned and operated by the State health authority and funded by the Commonwealth. This means that any action against the facility would be taken against the responsible state authority.

A growing body of High Court cases in the 1990's were seen to be finding in favour of defendants[31]. It was suggested that this was because judges were reluctant to make findings of negligence where such judgement would likely bankrupt the defendant[32], which would most likely be the case in actions against the generally non-union PCAs and any general-duties RNs. However, this trend was seen to be not as strong in the lower courts, especially where the defendant was a public authority[33]. Over the last two decades courts began to recognise the need for government and public authority defendants to be treated differently to all other defendants: for reasons of immunity[34]; budgetary consideration[35]; reliance on taxpayer funds[36]; and to ensure liberty of regulatory design which the courts felt should be free of a common law duty of care[37]. In the decade prior to Oakden and inspired by the 2002 *Ipp Review* into the Law of Negligence[38], tort reform legislation in most Australian jurisdictions began incorporating provisions creating policy defences[39] specifically directed at protecting government defendants from liability[40]. While the absence of such overt tort reform in the Northern Territory and South Australia may seem to some to make them softer targets, the tort reforms enacted in other states have generally only achieved re-statement of existing common law principles, potentially making the overall difference negligible.

---

[28] Department of Health, 'Primary mental health care services for people with severe mental illness' (2018) < https://bit.ly/2lEK4Ym>

[29] Parliament of the Commonwealth of Australia, 'Report on the inquiry into the Quality of Care in Aged Care Facilities in Australia' (2018, House of Representatives Standing Committee on Health, Aged Care and Sport), Canberra, ACT.

[30] *Ibid*.

[31] Hon. J. J. Spigelman, 'Negligence: The last outpost of the welfare state' (2002) 76 *The Australian Law Journal* 7, 432-451.

[32] *Ibid* at 433.

[33] *Ibid* at 434, citing as examples: *Crimmins v Stevedoring Industry Finance Committee* (1999) 200 CLR 1; and, *Brodie v Singleton SC* (2001) 75 ALJR 992.

[34] *Mansfield v Great Lakes Council* [2016] NSWCA 204: confirmed the strict thresholds in regards to reasonableness and knowledge that must be established before a government or public authority can be found negligent, and provides important analysis of provisions of the *Civil Liability Act 2002* (NSW) s43A and s45 that provide immunity from liability for authorities and their agents in the exercise of statutory powers.

[35] *Crimmins v Stevedoring Industry Finance Committee* (1999) 200 CLR 1, at 21 per Gaudron J; at 34 per McHugh J.

[36] F McGlone & A Stickley, *Australian Torts Law* (LexisNexis Butterworths, 2nd ed, 2009) at p 211.

[37] *Graham Barclay Oysters Pty Ltd v Ryan* (2002) 211 CLR 540 at 606-607.

[38] The Hon. David Ipp, 'Review of the Law of Negligence: Final report' (September 2002, Department of the Treasury) Canberra, ACT, <https://static.treasury.gov.au/uploads/sites/1/2017/06/R2002-001_Law_Neg_Final-2.pdf>

[39] J Bell-James & K Barker, 'Public Authority liability for negligence in the post-Ipp era: Sceptical reflections on the 'Policy Defence'' (2019) 40 *Melbourne University Law Review* 1.

[40] M Aronson, 'Government liability in negligence' (2008) 32 *Melbourne University Law Review* 44.

**The Negligence Case**

Australian courts currently apply a multi-factorial approach that involves weighing up various *salient features* that include: foreseeability, vulnerability of the plaintiff, control exercised and any assumptions of responsibility by the defendant[41]. This approach was unanimously endorsed by the Full Court of the High Court of Australia in *Sullivan v Moody*[42]. Use of the multi-factorial approach has allowed negligence law to incrementally develop through judicial examination of analogous cases and the application of relevant factors to the facts at hand[43]. Most states and territories require actions in negligence to be brought within six years of when the cause of action first arises[44], the exception being the Northern Territory which has a three-year limitation period[45].

**Bayesian Network for Case Outcome Determination**

Bayesian Networks (BNs) are built on Bayes' theorem and provide a graphical framework for compact representation that enable the decision-maker to reason probability under conditions of uncertainty[46]. BNs capture intuitive causal relationships and have been demonstrated as an effective decision-support tool in the legal context[47]. For example, they can be used to help determine causality[48], attribute responsibility for harm[49], and generally support the process of legal argumentation[50].

The BN model is comprised of two main components: *structure* and *parameters*. Determining the *structure* requires three steps: (i) identifying the variables important to your problem; (ii) defining each variable's states; and, (iii) identifying the relationships, or arcs, between variables and the direction of those relationships. *Parameters* are then applied to each variable to represent the strength of relations in the BN structure.

The initial process for developing a BN capable of reasoning the outcome of negligence claims required an audit of cases where a Government or statutory authority was defendant at first appearance. The audit, an extract of which is included in Appendix A, was used initially to identify variables that may be determinate in negligence cases, which are shown in the mind map in Figure 1.

---

[41] See *Caltex Refineries (Qld) Pty Ltd v Stavar* [2009] NSWCA 258, [103] for a general summary of the salient features.

[42] *Sullivan v Moody* [2001] HCA 59; 207 CLR 562; 75 ALJR 1570 at [50]: *Different classes of case give rise to different problems in determining the existence and nature or scope, of a duty of care*. Once this class of case and its various problems have been identified, at [50] *the relevant problem will then become the focus of attention in judicial evaluation of the factors which tend for or against a conclusion, to be arrived at as a matter of principle*.

[43] J Bell-James & A Huggins, 'Compliance with statutory directives and the negligence liability of public authorities: The case of climate change and coastal development', (2017) 34 *Environmental and Planning Law Journal* 5, pp 398-417.

[44] For example: *Limitations Act 1969* (NSW), s14; *Limitation of Actions Act 1974* (Qld), s 10(1)(a); *Limitation of Actions Act 1936* (SA), s 35(c).

[45] *Limitation Act 1981* (NT), s 12(1)(b).

[46] Norman Fenton & Martin Neil, *Risk assessment and decision analysis with Bayesian networks* (CRC Press, 2012).

[47] N Fenton, M Neil & D Berger, 'Bayes and the Law' (2016) 3 *Annual Review of Statistics and its Application*, 51-77.

[48] David Lagnado & Tobias Gerstenberg, 'Causation in Legal and Moral Reasoning' in *Oxford Handbook of Legal Reasoning* (Oxford Press, 2017) 565-602.

[49] H Chockler, N Fenton, J Keppens & D Lagnado, 'Causal analysis for attributing responsibility in legal cases' (paper presented at The 15th International Conference on Artificial Intelligence & Law, San Diego, USA, June 8-12 2015).

[50] F Taroni, C.G. Aitken, P Garbolino & A Biedermann, '*Bayesian networks and probabilistic inference in forensic science*' (Wiley, 2016) 372; and, Norman Fenton, Martin Neil & David Lagnado, 'A general structure for legal arguments about evidence using Bayesian networks' (2013) 37 *Cognitive Science* 1, 61-102.

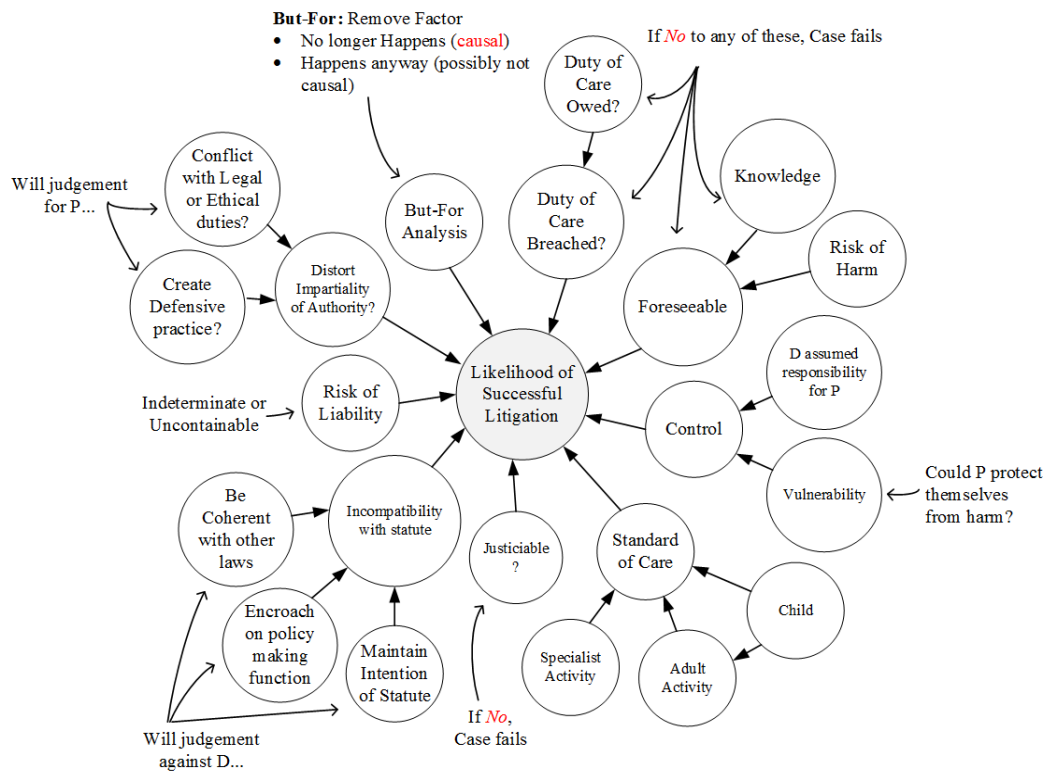

*Figure 1: Mindmap from Case Audit*

**The Simplest Model**

A common approach to designing BNs for legal reasoning is the story model[51]. The story model aids development of rich narrative-based explanations of evidence, and arises out of the assumption that jurors construct a story to make sense of the evidence in an attempt to fully support a verdict[52]. While the story model can be applied to explain a given case, the resulting model remains bespoke to that case and cannot always be generally applied. Nodes in a story model reflect the actions of the defendant and the justiciable question the trier of fact is tasked to resolve – for example: *did J stab C and cause C's death?*[53] In order to create the simplest model that could be more generally applied to cases where a public authority was defendant, it was necessary to focus the story model on statutory and common law elements: those factors that decisions in a negligence case are based on when the court finds in favour of either the plaintiff or the public authority defendant. The simplest model for negligence cases where a public authority is defendant was initially learned from the data captured in the case audit, and was further refined using expert knowledge. The parameters for each node except the Necessary requirements and case outcome nodes, which are synthetic nodes necessary to organise and simplify the model whose parameters are defined by simple logic, were learned from the extract data in Appendix A. The resulting BN model is presented in Figure 2.

---

[51] E.g.: 48 and 49.
[52] 48 at 575.
[53] 48 at 577.

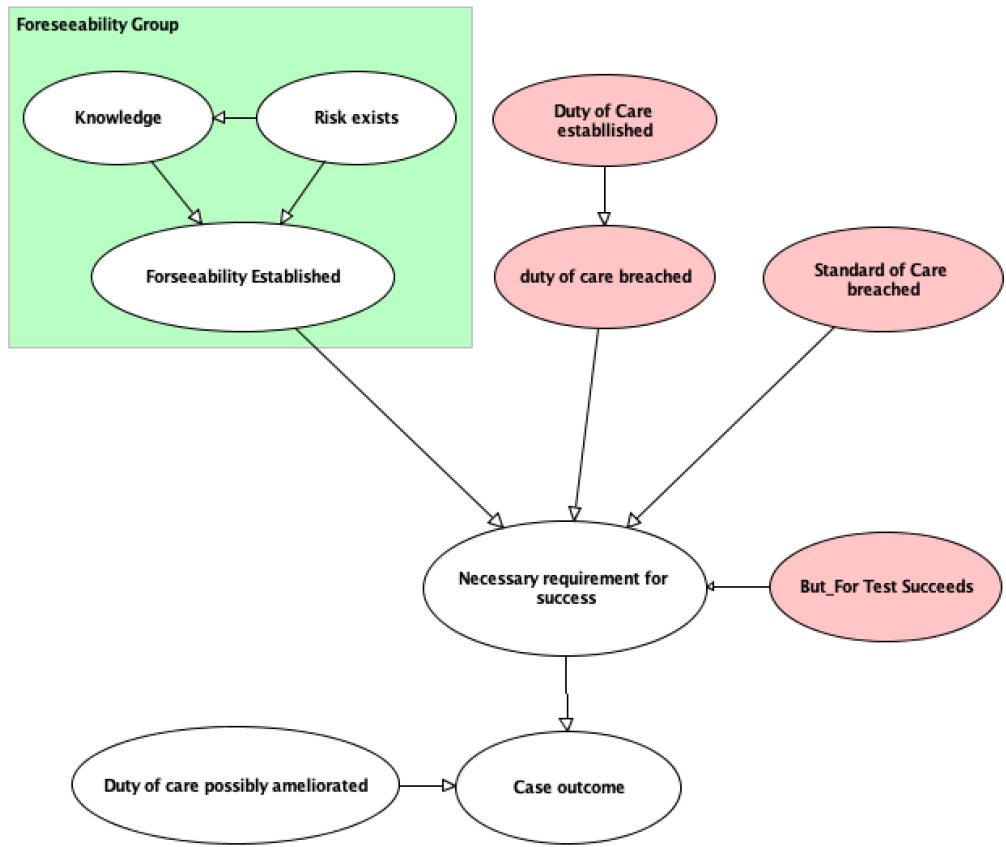

*Figure 2: The simplest BN Model*

To validate the model structure and parameters the Plaintiff Does Win backwards inference scenario was modelled. As Figure 3 shows, the conditions necessary for a successful plaintiff case include that three necessary requirement nodes must be true, including that: (i) *foreseeability is established*; (ii) the *duty of care is breached*; and, (iii) the *but-for test succeeds*. It was observed from the case audit data that in order for foreseeability to be established the plaintiff had to show a *risk of harm* existed and the public authority defendant had *knowledge* of that risk. This was demonstrated in the model in that both parent nodes for *foreseeability established* are absolutely true. Similarly, in order for a duty of care to be breached that duty had to have already been established. For this reason, the *Duty of Care established* node is also observed to be absolutely true.

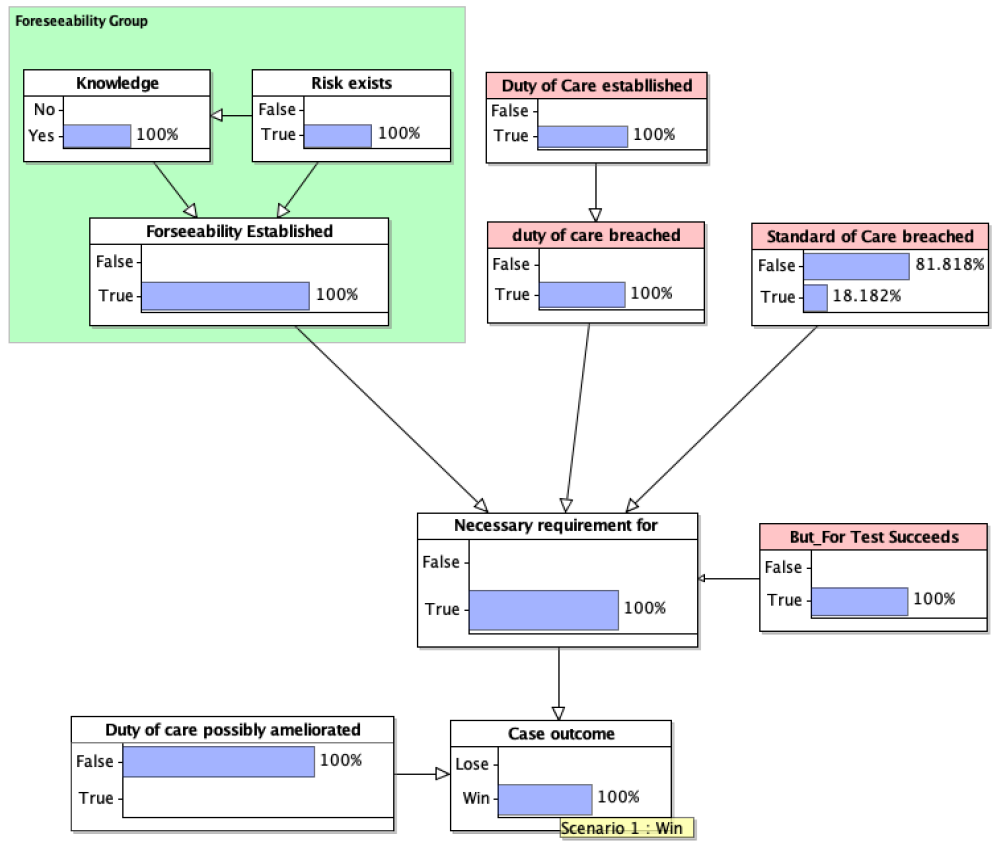

*Figure 3: The Plaintiff Does Win backwards inference scenario*

To evaluate whether successfully establishing the three necessary requirement nodes was sufficient for a plaintiff win, the *Plaintiff Should Win* scenario was modelled and is shown in Figure 4. As with the first scenario, observing *Foreseeability Established* as true causes *Risk exists* and *Knowledge* to also both be true. Similarly, observing *Duty of Care Breached* to be true results in *Duty of Care established* also being true. Where the BNs evaluation of the parameters learned from the case audit data differs is in the *Case outcome*: we now see that the plaintiff has only a 22% probable likelihood of achieving a win. What the *Plaintiff Should Win* scenario tells us is that in a case with a government or public authority defendant, where the plaintiff is able to establish foreseeability, that a duty of care was breached, and that but for the defendant's actions (or inactions) the harm caused to the plaintiff would not have resulted, the plaintiff still only has slightly better than a *one in five* chance of achieving victory and receiving a damages award. The primary factor in the much higher number of losses for plaintiffs are the amelioration factors: that government and public authorities are ever-increasingly able to avoid liability for breaching their duty of care through policy and statutory defences that allow them, amongst other things, to claim immunity.

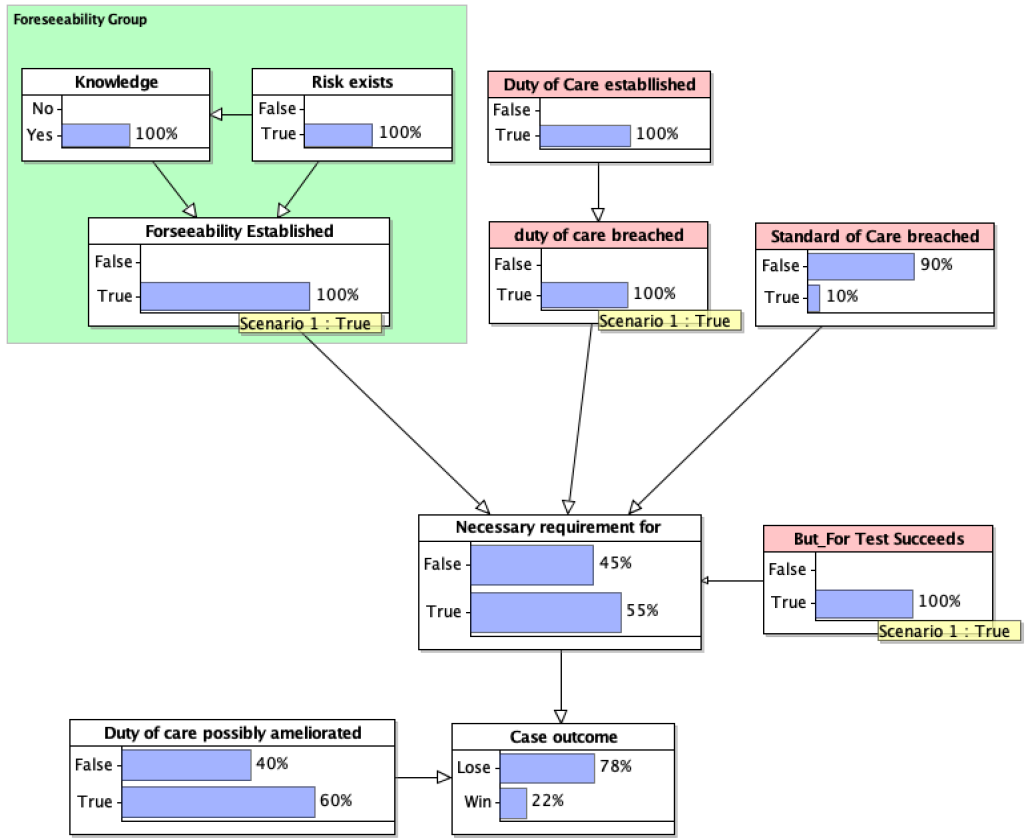

Figure 4: The Plaintiff Should Win scenario

**Discussion**

There are numerous strengths to the approach described in this paper. Rather than focusing on a story drawn from the evidence presented in a single case, and therefore being unique to that case, the presented model was established from caselaw drawn out of multiple similar cases and is therefore capable of general application to that type of case. Each was a negligence lawsuit against a public authority defendant, following the same adjudicative framework and requiring the plaintiff to demonstrate presence of the same elements for a successful outcome. The model is also capable of demonstrating something that lawyers may find surprising: that even when the plaintiff's case possesses the most necessary elements for success, in almost four out of five cases, the public authority is still more likely to be victorious.

A limitation of this work is the low number of cases that could be audited in the time available to provide data in the model's creation. Auditing cases necessitated intense review of entire judgements which, when a judgement can consist of *four-hundred or more* paragraphs, is time consuming. Where time and resource permitted the case was followed to its highest appearance in order to ensure that the most recent or final outcome was the one reported. In order to populate all elements that were part of the mindmap, and hence collected in the full case audit, it was necessary to read beyond majority findings and into minority or dissenting verdicts as well. There was potential for elements to be missed as this additional effort could become tedious: especially where points that had been made in the majority verdict were repeated, or where a judge laboured a particular point by comparing and contrasting the instant case against points of sometimes *twenty or more* prior reported cases.

**Conclusion**

This paper has reviewed the unfortunate and distressing situation that arose at Oakden. It identified those who could bring causes of action and the parties that could potentially be called to defend them. It was clear that more than any other, one defendant is in the financial position that makes it the most likely target for litigation: the government authorities that owned, operated and funded the facility. An audit of negligence cases where a government or public authority was defendant at first appearance was conducted and it, along with expert knowledge, was used to construct a Bayesian model capable of reasoning the necessary elements and case outcomes. Once trained with data gathered from those cases, the model was able to show that in a negligence matter against the government or a public authority the plaintiff who meets all of the necessary requirements to succeed is still more likely to lose than achieve the damages windfall they seek.

**About the Authors:**

**Scott McLachlan** has an MPhil in Information Science (Massey), a GDL (Waikato) and an LLM (ANU). Mr McLachlan recently completed his PhD with the Risk and Information Management (RIM) group at Queen Mary, University of London (QMUL) researching Bayesian-based decision solutions for health and legal applications.

**Evangelia Kyrimi** has a BSc in Statistics (Athens), MSc in Biostatistics (Descartes) and PhD in Statistics (QMUL). Ms Kyrimi is an experienced health and medical researcher and biostatistician and is currently a post-doctoral researcher and decision scientist with the RIM group at QMUL.

**Norman Fenton** has a BSc in Mathematics (Sheffield), MSc in Mathematics (Sheffield) and PhD in Mathematics (Sheffield). He is both Chartered Engineer and Mathematician, holds several fellowships and is a certified and experienced expert witness assisting the Court to understand issues of causal reasoning and probability. He is presently Professor of Computing and Director of the RIM group at QMUL.

# Appendix A: Extract from Case Audit

| Case | Won/Lost | DoC Est. | DoC Breach | SoC Breach | But/For | DoC Ameliorated | State | L/S/F | Authority Defence | Notes |
|---|---|---|---|---|---|---|---|---|---|---|
| Mansfield [2016] NSWCA 204 | Lost | Yes | No | No | Failed | Yes | NSW | L | Immunity from Liability under s 43A and 45 of Civil Liability Act 2002 (NSW) | 1. Evidence failed to establish any officer with requisite authority had actual *knowledge* of the risk for harm that materialised [3] making the claim *Inconsistent* with the statutory scheme [4] |
| Lane [2013] NSWDC 13 | Lost | Yes | No | No | Failed [102] | No | NSW | S | Absolute defence against all claims. That care provided was not negligent: it was adequate or exceeded adequacy. | 1. Even though it is against a Local Area Health network, this is part of state provider, NSW Health, and was treated the same when appearing as a D. |
| Vairy [2005] HCA 62 | Won | Yes [130] | Yes [199] | No | Succeeded [211] | No (Claimed but not successful) | NSW | L | That the plaintiff was incorrect as to salient facts (water level) [57] That having regard to the powers and responsibilities conferred on it by the Local Govt Act it did NOT owe a duty to warn P of obvious risks for injury [127] That erecting of a sign would be an imposition on the natural amenity of the location [165] Sought to claim contributory negligence but this arm failed [223] | 1. Council had care, *control* and management. 2. Council was aware [103] (*knowledge*) that people were jumping from the rocks and therefore the potential for harm that P suffered). Council officer with authority also had *knowledge* [107]. 3. Council accepted it *assumed responsibility* for those entering the reserve [127] 4. Harm was *foreseeable* [133] |
| Scarf [1998] QSC 233 | Lost | Yes [68] | No [76] | No | Failed [37 and 42] | Yes | QLD | S | State of QLD Defence that but/for test should fail – that even with a sign P would have done act. | 1. The court did not establish *knowledge* for the state or local authorities [24 and 25] 2. Risk was *foreseeable* [63] 3. Commissioner had *control* 4. Reasonable foreseeability of a real risk of injury can create special relationship 5. Practical considerations should balance out the DoC [72] |
| Scarf [1998] QSC 233 | Lost | Yes [68] | No [76] | No | Failed [37 and 42] | Yes | QLD | L | Council of City of Gold Coast Council claimed it had no authority (*control*) to erect sign [61] | 1. The court did not establish *knowledge* for the state or local authorities [24 and 25] |

| Case | Result | | | | | | | | Defence | Notes |
|---|---|---|---|---|---|---|---|---|---|---|
| Brodie [2001] HCA 29 | Lost | No [333, 371] | - | No | Failed | Yes | NSW | L | Immunity for non-feasance | 1. D not liable to rectify deteriorating roads in the shire<br>2. P drove over-weight truck over bridge in spite of sign warning to contrary |
| Shirt [1980] HCA 12 | Won | Yes [10] | Yes [18] | No | Succeeded [21 and 23] | No | NSW | L | That the causation should fail because P was aware of rocks where he was diving. Court said that but for a sign P would likely not have jumped. | 1. Generalised duty of care requires authority to take reasonable steps to avoid *foreseeable* risk of injury<br>2. *Foreseeability* found at [12]<br>3. Trial judge considered proximity and found Authority had *assumed responsibility* in relation to persons attending location [7]<br>4. Board had full *knowledge* of topography and therefore risk [Brennan at 4] |
| Barclay [2002] HCA 54; 211 CLR 540 | Lost | No [33] | - | No | - | No | NSW | S | | 1. Allegations against state do not allege carelessness in exercise of a statutory power [5]<br>2. The negligence claims invite judiciary to judge reasonableness of conduct of legislative or executive arms of govt. [6]<br>3. Claim based on non-feasance [8]<br>4. Finding DoC against govt requires finding un/reasonableness of its actions [15]<br>5. State said to have managerial *control* over fisheries [34]<br>6. Power to protect public does not give rise to duty of care to protect individual [32] |
| Barclay [2002] HCA 54; 211 CLR 540 | Lost | No [40] | - | No | - | Yes (see 5) | NSW | L | | 1. Allegations against council do not allege carelessness in exercise of a statutory power [5]<br>2. Claim based on non-feasance [8]<br>3. Council had no managerial *control* over fisheries [34]<br>4. Power of regulation over an area or activity vested by statute does not mean council owes DoC to individual over issues of exercising that power [35] |

| Case | Result | | | | | | State | L/S | Defendant / Arguments | Reasoning |
|---|---|---|---|---|---|---|---|---|---|---|
| | | | | | | | | | | 5. Powers for benefit of public generally and not for individual or specific class [39] (*intention*) |
| Romeo (1998) 192 CLR 431 | Lost | Yes [54, 56, 141] | - No [54, 56] | No [166] | Failed [132] | Yes | NT | S | That being a govt authority exempted liability due to financial and political reasons – termed *policy factors* [138] – failed or was not fully considered at [141] | 1. While one judge says no DoC, two others say there is a general DoC but that it only required P to do what was reasonable – fencing of a 2km cliff not reasonable<br>2. Risk not reasonably *foreseeable* [160-163]<br>3. The main but/for argument seems to be two-fold: the claim that but for a sign or fence failed, while the counter claim that but for her intoxication she might have taken more reasonable steps to protect herself. |
| Ballerini [2005] VSCA 159 | Won | Yes [3, 10, 26] | Yes [3, 11] | Yes | Succeeded<br>Considered at [4, 68, 71] | No | VIC | L | Berrigan Shire Council<br>Contributory Negligence – reduction in damages of 30% [6]<br>Council claimed to owe no duty of care and no control of site or log. [23]<br>Pointed finger for liability re log at commission alone [23, 34]<br>Claimed that judge erred beyond extent of reasonableness [38] | 1. Standard of care: Reasonable Care<br>2. Council had requisite *knowledge* that activities like that of the P occurred there [10 & 26]<br>3. Reasonably *foreseeable* risk [10, 28]<br>4. Council had *control* of land [26] |
| Ballerini [2005] VSCA 159 | Lost | Yes [5] | No [5, 14] | No [5] | - | Yes | VIC | S | State Forestry Commission<br>Commission had different charter and purpose [14]<br>Had no relationship to P (proximity) [14] | 1. Authority only had peripheral responsibility [6] |
| Dederer [2006] NSWCA 101 | Lost | Yes [16] | No [21, 33] | No | Failed | Yes | NSW | L | Great Lakes Shire Council<br>Result of the materialisation of an obvious risk of a dangerous recreational activity" within the meaning of s 5L of the *Civil Liability Act 2002* (NSW) [26]<br>That s310 of Legal Government Act does not impose statutory duty of care [12]<br>They also claimed that as they were never notified that construction was completed, the statutory duty under cl83 of Ordinance 71 was never invoked [12] | 1. Council had *knowledge* of the behaviour/activity that resulted in harm<br>2. Council assumed duty (*control*) [8]<br>3. Statute did not impose a duty on the council [12]<br>4. Duty mitigated by policy [21] |
| Dederer [2007] HCA 42 | Lost | Yes [43] | No [78, 80] | No [55 onwards] | Failed | Yes | NSW | S | Roads and Traffic Authority | 1. RTA responsible for construction and management (*control*) [2] |

| Case | | | | | | | | | | | |
|---|---|---|---|---|---|---|---|---|---|---|---|
| | | | | | | | | | | | 2. Potential P's should exercise reasonable care for their own safety [45]
3. DoC was to all users of bridge, not P in particular (*intention*) [47]
4. This appeal recognised that the but for test in the prior hearing had incorrectly said that the RTA knowing people ignored the signs should have done more. This was found incompatible with the law and application hence RTA won. RTA had met reasonableness test by installing signs and keeping road safe |
| Heyman [1985] HCA 41 | Lost | No [44] | - | - | Failed | No | NSW | L | | | 1. Foreseeability of P's reliance means it is positive conduct on the part of the Authority that attracts the DoC [28]
2. Council lacked *foreseeability*
3. No statutory duty imposed; therefore, any duty would have been reasonableness
4. Damage suffered had to be caused by negligent failure to act of the council. Given the council had no duty TO act, their inaction was not the cause of the damage. Rather it was caused by poor building quality of third party |
| **TOTALS** | **Won:** 3 **Lost:** 12 | **Yes:** 11 **No:** 4 | **Yes:** 3 **No:** 12 | **Yes:** 1 **No:** 14 | **Failed:** 9 **Succeeded:** 3 **Not Cons:** 3 | **Yes:** 9 **No:** 6 | **NSW:** 10 **VIC:** 2 **QLD:** 2 **NT:** 1 | **L:** 9 **S:** 6 **F:** 0 | | | |